\documentclass[3p,times]{elsarticle}
\usepackage{graphicx}
\usepackage{textcomp}
\usepackage{graphicx}
\usepackage{epstopdf}
\usepackage{subfigure}
\usepackage{color}
\usepackage{float}
\usepackage{ntheorem}
\usepackage{enumerate}
\usepackage{ecrc}
\usepackage{times}
\usepackage{amsmath,amssymb}
\usepackage{ntheorem}
\newtheorem{rem}{Remark}
\volume{1}

\firstpage{1}

\journalname{Preprint submitted to XXXX}

\runauth{sss}

\jnltitlelogo{Version 0.2}

\usepackage{amssymb}
\usepackage[colorlinks,
            linkcolor=red,
            anchorcolor=blue,
            citecolor=green
            ]{hyperref}
\def\sectionautorefname~#1\null{Section~#1\null}
\def\subsectionautorefname~#1\null{Section~#1\null}
\def\subsubsectionautorefname~#1\null{Section~#1\null}
\def\figureautorefname~#1\null{Fig.~#1\null}
\def\tableautorefname~#1\null{Tab.~#1\null}
\def\equationautorefname~#1\null{Eq.~(#1)\null}

\usepackage[figure]{hypcap}
\biboptions{sort&compress}
\usepackage[ruled]{algorithm2e}
\usepackage[figuresright]{rotating}

\begin{document}
\begin{frontmatter}



\dochead{}

\title{\LARGE \textcolor[rgb]{0,0,1}{Effect of density control   in partially observable asymmetric-exit  evacuation under guidance}: Strategic suggestion under  time delay}


%\author[Ren and Gao]{Huan Ren}
%\author[Yan]{Yuyue Yan\corref{mycorrespondingauthor}}\cortext[mycorrespondingauthor]{Corresponding author.}\ead{yan.y.ac@m.titech.ac.jp}    % Add the
%\author[Ren and Gao]{Fengqiang Gao}            % e-mail address
\author[Gao,Chen]{Fengqiang Gao}
\author[Yan]{Yuyue Yan\corref{mycorrespondingauthor}}\cortext[mycorrespondingauthor]{Corresponding author.}\ead{yan.y.ac@m.titech.ac.jp}    % Add the
 \author[Chen]{Zhihao Chen}
 \author[Zheng]{Linxiao Zheng}
 \author[Chen]{Huan Ren}
           % e-mail address
  % (ead) as shown
\address[Gao]{School of Aerospace Engineering, Xiamen University, Fujian 361102, China}
\address[Chen]{School of Information Science Technology, Xiamen University Tan Kah Kee College, Fujian 363105, China}
\address[Yan]{Department of Systems and Control Engineering, Tokyo Institute of Technology, Tokyo 152-8552, Japan}  % Please supply
\address[Zheng]{College of Information Science and Engineering, Huaqiao University, Fujian 362021, China}
\begin{abstract}
To enhance the evacuation efficiency in partially observable asymmetric-exit evacuation \textcolor[rgb]{0,0,1}{under guidance, a general framework of the dynamic guiding assistant system
is presented to investigate the  effect of density control.}
In  this framework,
several evacuation assistants  are established to observe the partial information of pedestrians'
location and adjust the guiding signals of the dynamic guiding assistant systems.
A  simple on-off-based density control algorithm  is proposed for the   evacuation assistants according to   the delayed data of the observed information (i.e., pedestrian densities in the observed regions near the corresponding exits).
This paper provides  strategic suggestions on how to set the observed region and the target density by involving a force-driven cellular automaton model.
It is observed that the proposed density control algorithm can control (positively affect) the global distribution of the pedestrians' locations and suppress arching phenomena in the evacuation process even using the  observed partial information under time delays.
By imposing a  moderate target density,  the dynamic guiding assistant system also suppresses the triggers of collisions around the exits and avoids  {inefficiently separating the pedestrians}. To enhance  evacuation efficiency, we reveal an interesting fact without loss of generality that  we only need to observe
the pedestrians' location from a small region near the exit instead of a large region when the time  delay of the observed information is  slight enough.  Our numerical findings are expected to provide  new insights \textcolor[rgb]{0,0,1}{into}  designing  computer-aided guiding strategies in  real evacuations.
\end{abstract}


\end{frontmatter}

% \linenumbers for ion

%% main textwhen the earthquake, fire, terrorist attacks, and other emergencies occur in public places
\section{Introduction}
In the last decades, the efficiency estimation problem has been  a key aspect \textcolor[rgb]{0,0,1}{of} evaluating safety performance for pedestrian evacuations.
 During the occurred emergency events (such as fire, earthquake, and terrorist attacks),  since the  pedestrians  in the evacuation space are usually myopic and many irrational behaviors may inherently and negatively affect  the pedestrians' decision making (which may eventually lead to an efficiency loss causing massive  casualties) in the evacuation process \cite{meng2019pedestrian,cheng2018emergence,haghani2019panic,guan2019towards}, some evacuation strategies are often beforehand designed to enhance the evacuation efficiency. In the literature, those evacuation strategies may include the choices on room layout \cite{liao2014layout}, number of exits \cite{liu2020fuzzy}, location of obstacles \cite{varas2007cellular}, to name but a few. To effectively evaluate  evacuation efficiency,
establishing  mathematical models  {to describe}  the pedestrians' {decision-making} behavior and the  resulting phenomena in the evacuation process should  be of prime importance \cite{lovaas1994modeling,helbing2000simulating,muller2014study,PhysRevE,huang2017behavior,alizadeh2011dynamic,fu2015influence,shahhoseini2019pedestrian,yang2021effect,chraibi2010generalized,hao2014exit,HRABAK2017486}.
%As the typical discrete-time dynamics, the cellular automaton models provide a structure for reflecting the pedestrians' microscopic characteristics
%with some simple update laws (evolution rules), which may be combined with the notion of social force \cite{chen2012study}, floor field \cite{varas2007cellular,li2019extended}, etc.
%In the literature \cite{lovaas1994modeling,helbing2000simulating,muller2014study}, the existing pedestrian dynamics are mainly categorized to the continuous-time and discrete-time models \cite{PhysRevE,varas2007cellular,huang2017behavior,alizadeh2011dynamic,cao2018exit,shahhoseini2019pedestrian,huang2015behavioral,chraibi2010generalized,liao2014layout,hao2014exit,HRABAK2017486}.
For example, a cellular automaton (CA) model describing the change of cooperative behavior in heterogeneous pedestrians was investigated in \cite{huang2017weighted} to
examine how the pedestrians' dependency relationship  \textcolor[rgb]{0,0,1}{influences}  evacuation efficiency.
A floor field model investigating agitated behaviors \textcolor[rgb]{0,0,1}{with} elastic characteristics was presented in \cite{xie2012agitated}, whereas  a fine discrete field CA model integrating anisotropy, heterogeneity, and time-dependent characteristics, was discussed in \cite{fu2018fine}.
%Alizadeh investigated a dynamic cellular automaton model for a evacuation space with obstacles and discussed the influences of doors position and width, pedestrians distribution on the evacuation efficiency \cite{alizadeh2011dynamic}.
A multi-grid CA model capturing the turning behaviors was proposed in \cite{miyagawa2020cellular}.
Furthermore, an extended floor-field CA model describing the group behaviors in crowd evacuation was found in \cite{lu2017study}, whereas
a modified social force model and a  force-driven CA model were proposed in \cite{yang2014guided} and \cite{REN2021125965} to reflect the evacuation process under guidance.% The herding behaviors under  view-limited conditions was studied in \cite{meng2019pedestrian}.

With the development of modern technology, especially in the aspect of \textcolor[rgb]{0,0,1}{the} imaging process, communication networks, and IoT technologies,
some evacuation assistants (or equipment) can be adopted in the (computer-aided) evacuation process \textcolor[rgb]{0,0,1}{to provide} accurate route
or exit information to the evacuees and hence help (guide) them to escape as efficiently as possible.
In the related works with   guiding behaviors \cite{yang2016necessity,ma2016effective,ma2017dual,zhou2019guided,chu2017variable,zhou2019optimization,long2020simulation,yang2020guide}, Yang et al. \cite{yang2016necessity} are the first  {researchers}  \textcolor[rgb]{0,0,1}{to claim} the necessity of using guides (dynamic leaders) in the evacuation process, whereas the effects of leadership in single-exit rooms were investigated in \cite{ma2016effective} with limited visible range.
The authors in \cite{ma2016effective} further  stated that  guiding behavior  may yield a negative effect when too many  pedestrians are neighbor to the leader.  Considering the situation where the dynamic leader attracts the crowds and  \textcolor[rgb]{0,0,1}{moves}  together with them towards one of the exits, the optimal number and  positions for those leaders were derived in \cite{yang2020guide} and \cite{yang2020pedestrian}.
Even though the literature  \cite{ma2016effective} and \cite{yang2020guide} have claimed that the density of   pedestrians plays an important  role in the evacuation process, to our best knowledge, our previous study in \cite{REN2021125965} is the first work  indicating the possibility and goodness \textcolor[rgb]{0,0,1}{of} controlling the pedestrians' density.  It was found that for a symmetric-exit evacuation space,  unbalanced pedestrian distributions may be yielded because of
the mutual attractions among the pedestrians, but  {those unbalanced distributions} can be suppressed by imposing some control mechanisms based on the data of pedestrian density from an observed region. Given the constant observed region, by adjusting the guiding signals according to the pedestrian densities around the exits, \cite{REN2021125965} showed  preliminary evidence that imposing control mechanisms may be able to pursue a more balanced pedestrian distribution and an efficient evacuation process.
However, the effects of the size of the observed regions are not yet revealed. Furthermore, it is essential  to note that the results in \cite{REN2021125965} did not consider the influence of \textcolor[rgb]{0,0,1}{data} time delay in the control mechanisms, which is unnatural for a realistic computer-aided evacuation process.

Contributions: Different from \cite{REN2021125965} and the other existing results,  the main contribution of  this paper is focused on finding the optimal  size of the observed regions under different data time delays for a dynamic guiding
assistant system in  partially observable asymmetric-exit evacuation. To this end, inspired by control theory, we propose a general framework of dynamic guiding
assistant systems with a density control algorithm by assuming that the evacuation assistants can observe only a few parts of the evacuation spaces.
We consider a simple on-off-based density control algorithm for   evacuation assistants with    delayed data of the
observed information, i.e., pedestrian densities in the  regions near the corresponding exits. By involving a force-driven
cellular automaton model, we discuss  strategic suggestions on  \textcolor[rgb]{0,0,1}{setting} the observed region  in this paper by presenting the numerical simulation results under  homogeneous and heterogeneous pedestrians with respect to \textcolor[rgb]{0,0,1}{the size of the visual field.} Our numerical findings contribute to \textcolor[rgb]{0,0,1}{providing} some new insights on
designing   computer-aided (control-based) guiding strategies in  actual  evacuations.

%\cite{ma2016effective,ma2017dual,li2017driving,guan2020cooperative,li2020relationship,cao2018exit,huang2017behavior,zhou2018optimal,zhou2019cellular,kurdi2020balanced,zhou2019guided,zhou2019optimization,li2019extended,yang2020guide,long2020simulation,ZHANG2021488}\cite{yue2011simulation2}.

This paper is organized as follows: A  framework of dynamic guiding
assistant system with density control algorithm is proposed in \autoref{section2}. In \autoref{section3}, we first introduce the  necessity of using dynamic guiding assistance  in the
asymmetric-exit evacuation process and then investigate the optimal size of the observed regions for various scenarios with a slight  data delay and a large  data delay. Finally, the conclusion is given in \autoref{section:4}.

\emph{Notation:} We use $\mathbb R_+$ for the set of positive real numbers, $\mathbb N_+$ for the positive integers,  and $\emptyset$ for the empty set.

\vspace{-10pt}

\section{System Description and Problem Formulation}\label{section2}
\vspace{-3pt}
\subsection{System Description}\label{Sec22}
Consider an evacuation system $\Pi$ consisting of several pedestrians in a multi-exits evacuation space. Each pedestrian dynamically changes its location over time  to escape from the evacuation space through one of the $m\in\mathbb Z_+$ number of exits according to its personal information, e.g., visible exit, communication with the other evacuees. We denote the set of exits by $\mathcal N_{\rm ex}\triangleq \{1,2,\ldots,m\}$. %and the set of remaining pedestrians at the time instant $t=1,2,\ldots,{+\infty}$ by $\mathbb {RM}_{t}$.%
As myopic decision-makers with limited information, the pedestrians may move in the same direction even when congestions exist around the target exit.
 In such a case,  the  exits may not be used efficiently. To help   pedestrians better efficiently escape from   space, in this paper, we consider a guiding  assistant system for the aforementioned evacuation system with $\varepsilon\in\mathbb N_+$ evacuation assistants (EAs), where the set of EAs is given by $\mathcal N_{\rm EA}\triangleq \{1,2,\ldots,\varepsilon\}$. In this system, each evacuation assistant $i\in\mathcal N_{\rm EA}$ is assigned to a corresponding exit $j\in\mathcal N_{\rm ex}$, and sends a \emph{guiding signal} to the pedestrians  {to attract}   them towards its own assigned exit. We assume that the guiding signal is dynamically adjustable by the evacuation assistant according to the \textcolor[rgb]{0,0,1}{real-time} situation of the evacuation process. Henceforth, we denote the guiding assistant system by $\mathcal G$.
\begin{figure}
  \centering
  \includegraphics[width=100mm]{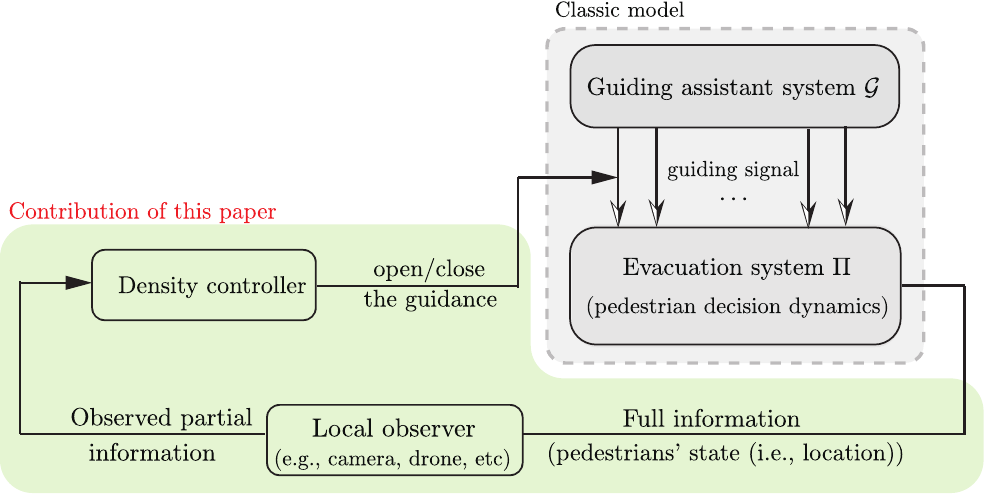}
  \caption{Structure of the guiding assistant-based evacuation system. Each evacuation assistant observes the partial information of agents' location  and adjusts the guiding signal according to the observed information. \textcolor[rgb]{0,0,1}{Involving the density controller, the classic (static) guiding assistant system becomes a dynamic guiding assistant system} where the guiding signals of $\mathcal G$ are understood as the inputs (feedback) to the evacuation system $\Pi$.  With a well-designed control algorithm (e.g., density control law),  the states (location) of the pedestrians and the evacuation efficiency may be controlled (positively affected). The detailed  control algorithm  and pedestrian evacuation  model considered in this paper are given in Sections~\ref{model} and \ref{Sec23} below.   }
  \label{fig:3}
\end{figure}

 {According to the control theory,} it is essential  to notice that the guiding signal of the guiding assistant system $\mathcal G$ can be understood as the \emph{input} to the evacuation system  so that the states (location) of the pedestrians are controlled (affected) by the EAs.
 This fact is characterized independently of the underlying mathematical models that describe the decision-making process of   pedestrians under guidance (guiding signals).
 However, as a common issue in control systems, the states (location) of the pedestrians may not be fully observed. Inspired by the classic control theory, we note that an unobservable system may \textcolor[rgb]{0,0,1}{still be} controllable under some well-designed control algorithms \cite{LIU201510,wang2018event}. Consequently, in this paper, we assume that EAs may  observe   partial information on the locations of the pedestrians (e.g., the pedestrian density $\rho_i$ near  exit $i\in\mathcal N_{\rm ex}$) using some  computer-aided technologies. For example, those EAs can observe the regions near   exits and  adjust the intensity of guiding signals according to the observed  information. A fundamental question is how to   set the size of the observed regions and  control the guiding signals to enhance the evacuation efficiency as much as possible.  To address this problem, we introduce a force-driven CA model under density control algorithm in the next section to implement the structure of the guiding assistant-based evacuation system
 illustrated in \autoref{fig:3}.

\begin{figure}
  \centering
  \includegraphics[width=100mm]{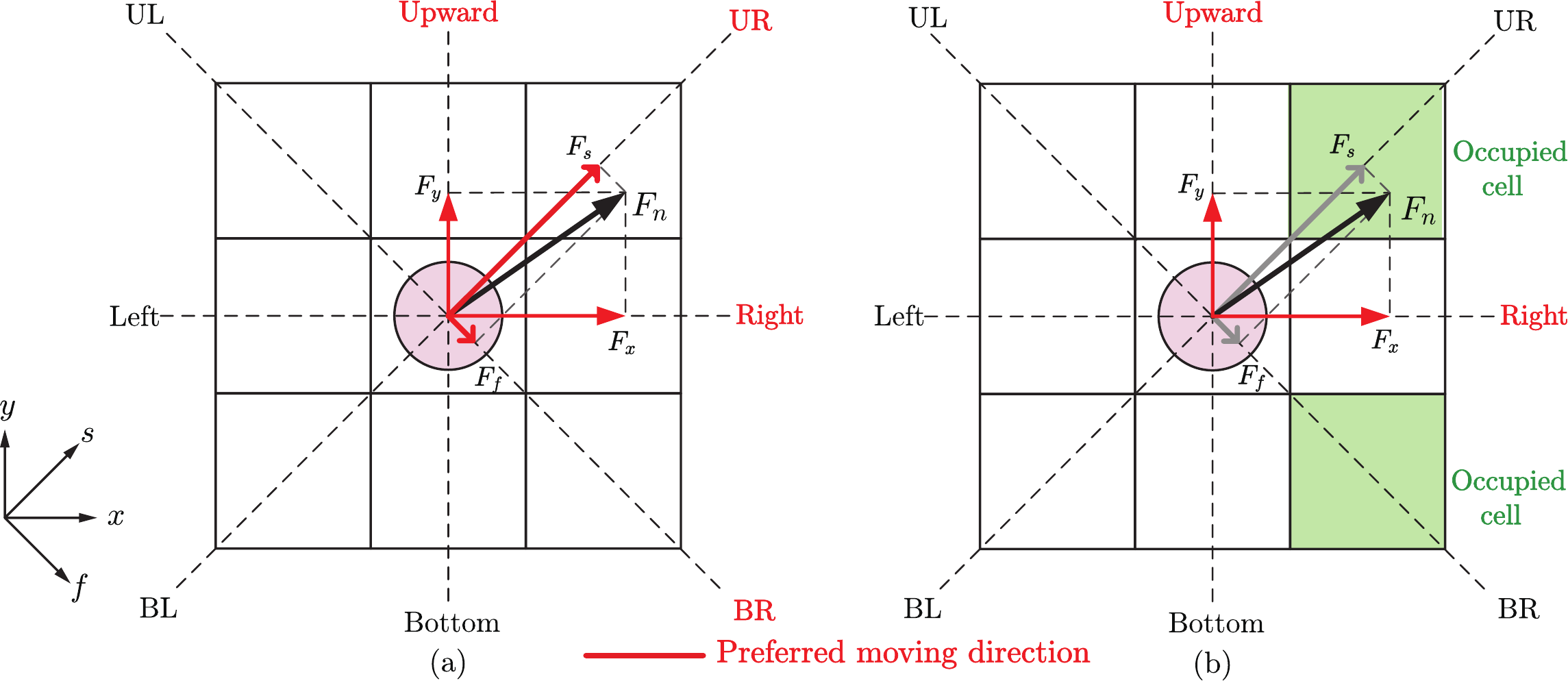}
  \caption{Preferred moving directions for a given social force $F_n$. In (a), the sequence of preferred moving directions is given by upper right, right, upward and bottom right.  In (b), as the cells in the upper right and bottom right are occupied, the sequence of preferred moving directions is given by  right and upward, i.e., $P_n^3=P_n^4=\emptyset$. Supposing that the pedestrians are only allowed to compete for 2 rounds in terms of  preferred moving directions (i.e., $k=2$), it is understood that the pedestrian $n$ in (a) (resp., in (b)) first competes for the upper right direction (resp., the  right direction). If he loses in the competition, then the next competition target is the right direction (resp., upward  direction) in (a) (resp., in (b)).}
  \label{fig:2}
\end{figure}
\vspace{-5pt}

%In the following section, we consider a force-driven CA model to describe the pedestrians' dynamic decision making in the evacuation space. such as social force models \cite{ma2016effective,ma2017dual} and cellular automaton model \cite{REN2021125965}
\subsection{Brief Introduction of Force-driven CA Model Under Density Control}\label{model}
In this section, we characterize the force-driven CA model to describe the guided evacuation process with the two-dimensional Moore  cells,  where each cell may be occupied by only one of the pedestrians with eight neighbor cells as the possible  moving directions for the next time instant (see \autoref{fig:2}(a)).
Specifically, \textcolor[rgb]{0,0,1}{to be consistent with the literature in guided
evacuation \cite{yang2016necessity,ma2016effective,ma2017dual,yang2020guide,yang2020pedestrian},}  we consider a square evacuation space and divide the space to $N\times N$ number of cells.
Letting  $\mathbb {RM}(t)$ denote the set of the pedestrians remaining in the evacuation system $\Pi$ at  time instant $t=0,1,2,\ldots$, we suppose that the remaining pedestrians simultaneously update their locations according to a social force-based evolution rule given in the following sections.

%For simplicity, in the rest of the paper, we assume that the number of exits and EAs are the same, i.e., $m=\varepsilon$. Furthermore, we assume that each of the EAs is located at its assigned exit.
 \vspace{-7pt}
\subsubsection{Force Definitions}
Before we present   the detailed expression of the evolution rule,   we define the social force consisting of a guiding force from an evacuation assistant, the mutual forces among visible pedestrians, and the attractive forces to  visible exits for each pedestrian $n\in\mathbb {RM}(t)$, \textcolor[rgb]{0,0,1}{i.e.,}
{\setlength\abovedisplayskip{4pt}
\setlength\belowdisplayskip{4pt}
\begin{align}\label{social}
    {F_{n} =    \textcolor[rgb]{0,0,1}{ w_1 F_{\rm guide}^n+ w_2\!\sum\nolimits_{m\in\mathbb {VP}_n}{F_{\rm mutual}^{n,m}}+ w_3\!\sum\nolimits_{i\in\mathbb {VE}_n} F_{\rm exit}^{n,i},}}
\end{align}
where} \textcolor[rgb]{0,0,1}{$w_1$, $w_2$ and $w_3\in\mathbb R_+$ denote the positive weighting factors depending on the evacuation scenario,} and $\mathbb {VP}_n$  (resp., $\mathbb {VE}_n$) represents the set of visible pedestrians  (resp., visible exits) for the pedestrian $n\in\mathbb {RM}(t)$.

Note that in \autoref{social}, $ F_{\rm guide}^n$ represents the guiding force pointing to the exit whose EA's guiding signal possesses a maximal \emph{signal to interference ratio},  i.e.,
{\setlength\abovedisplayskip{4pt}
\setlength\belowdisplayskip{4pt}\begin{equation}\label{guiding}
    F_{\rm guide}^n = u_{I_n}D  \vec r_{n,I_n},\quad    I_n\triangleq\arg\max_{i\in\mathcal N_{\rm EA}}\left(\frac{u_i}{1+\sum_{j\in\mathcal N_{\rm EA}\setminus\{i\}}(r_{n,i}^2/r_{n,j}^2)}\right),
\end{equation}
where} $u_{i}\in[0,1]\subset\mathbb R$ refers to the intensity coefficient controlled by the evacuation assistant $i\in\mathcal N_{\rm EA}$, $D\in\mathbb R _+$ refers to  the maximal  intensity of the guiding signal, $r_{n,i}$ refers to the distance (steps) to the exit  $i$,
and $\vec{r}_{n,I_n}$ refers to the unit vector pointing from \textcolor[rgb]{0,0,1}{the} pedestrian $n$ to the  exit $I_n$ (see Ref.~\cite{REN2021125965});
$F_{\rm mutual}^{n,m}$ represents a mutual force  from the visible pedestrian $m$  \cite{chen2012study,REN2021125965}, i.e.,
{\setlength\abovedisplayskip{4pt}
\setlength\belowdisplayskip{4pt}\begin{eqnarray}\label{mutual}
F_{\rm mutual}^{n,m}= \begin{cases} \textcolor[rgb]{0,0,1}{{\eta_1}\vec{r}_{m,n}},& \mbox{if $r_{n,m}=1$} \\
  {\eta_2}/{r_{n,m}^2}\vec{r}_{n,m}, &\mbox{if $1<r_{n,m}\leq w$}\end{cases}, \quad m\in \mathbb {VP}_n,
 \end{eqnarray}
where}  $\eta_1\in\mathbb R_+$ refers to the repulsion coefficient, $\eta_2\in\mathbb R_+$ refers to the attraction coefficient, $w$ refers to the length of the \textcolor[rgb]{0,0,1}{visual field},  ${r}_{n,m}$ refers to the distance (steps)  between   the pedestrians $m$ and $n$, and $\vec{r}_{n,m}$ refers to the unit vector pointing to the pedestrian $m$;
%Therefore, the interaction forces consider the cases of attraction and repulsion and inefficacy. Taking reference of the coulomb force, the pedestrian in different distance are assumed to be the like charges or unlike c harges, and the quantity of electricity is asserted to be 1.0 Unit.$^{12,14}$  Thus,
$F_{\rm exit}^{n,i}$ represents an attractive force to a visible exit \cite{chen2012study,REN2021125965}, i.e.,
{\setlength\abovedisplayskip{4pt}
\setlength\belowdisplayskip{4pt}\begin{equation}\label{exit}
    F_{\rm exit}^{n,i}=  E\vec{r}_{n,i}, \quad i\in\mathbb {VE}_n\triangleq\{i\in\mathcal N_{\rm ex}:  r_{n,i}\leq w\},
\end{equation}
where}  $E\in\mathbb R_+$ refers to the constant intensity, $r_{n,i}$ and $\vec{r}_{n,i}$ respectively refer to the distance and the unit
vector to the exit $i\in\mathcal N_{\rm ex}$.

\vspace{-3pt}
\begin{rem}\emph{
\textcolor[rgb]{0,0,1}{It is necessary  to notice that the repulsions between the pedestrian and obstacle may need to be further added
in the expression of the social force (\ref{social}) when the obstacle appears in the visual field. The expression of repulsions between the pedestrian and obstacle
can be found in \cite{chen2012study}. To be consistent with the literature in guided evacuation (see \cite{yang2016necessity,ma2016effective,ma2017dual,yang2020guide,yang2020pedestrian}), we do not consider an evacuation space with any  obstacle and hence those  forces are excluded in this paper.  On the other hand, when the heterogeneity
on quality of the pedestrians \cite{chen2012study} is considered, the expressions of the repulsions and attractions
should be changed to $
F_{\rm mutual}^{n,m}=
\begin{cases}  {\eta_1}q_mq_n\vec{r}_{m,n},& \mbox{if $r_{n,m}=1$} \\
  {\eta_2}q_mq_n/{r_{n,m}^2}\vec{r}_{n,m}, &\mbox{if $1<r_{n,m}\leq w$}
  \end{cases} $,
   where $q_m$ is the quality of the pedestrian $m\in \mathbb {VP}_n$ and $q_n$ is the  quality of the pedestrian $n$.}}
 \end{rem}\vspace{-6pt}

 After defining the notion of social forces,  the preferences of moving direction
candidates  are consequently derived by sorting the magnitude of the decomposed forces of \emph{F}$_{n}$ at \emph{x}, \emph{y}, \emph{s}, and \emph{f}-axis (see \autoref{fig:2}). Henceforth, we denote the sequence of ordered preference target cells  by $P_n=[P_n^1,P_n^2,P_n^3,P_n^4]$, where $P_n^i$ refers to the $i$-th preferred target.

\vspace{-7pt}
\subsubsection{Social Force-Based Evolution Rule}

Given an initial distribution of the pedestrians $\mathbb {RM}(0)$ and the intensity coefficients $u_i(0)$, $i\in\mathcal N_{\rm EA}$ of the guiding signals  from the guiding assistant system $\mathcal G$, the social force-based evolution rule of the force-driven CA model is  summarized as follows:
\vspace{-6pt}
\begin{enumerate}[(i)]\setlength{\itemsep}{-0.08cm}
\item Calculate the social force according to \autoref{social} and determine the sequence of ordered preferred target cells  $P_n=[P_n^1,P_n^2,P_n^3,P_n^4]$  for all the remaining pedestrians  $n\in\mathbb {RM}(t)$. If some of the desired targets have already been  occupied, then some elements of $P_n$ are understood as empty. For example,  the  cells  in the upper right and bottom right
directions  in \autoref{fig:2}(b) are not considered as the  preferred target cells and hence $P_n^3=P_n^4=\emptyset$ because those cells have \textcolor[rgb]{0,0,1}{already been} occupied.
\item Collisions and target competition. We suppose that each pedestrian $n\in\mathbb {RM}(t)$ possesses at most $k$ chances  {to  compete for} the target cells with the other pedestrians according to  its own sequence $P_n$.  If a cell is targeted as the preferred moving direction by several pedestrians in one of the competition rounds, it turns out that \emph{collisions} may happen \cite{tanimoto2010study} so that none of the conflicting pedestrians wins in the competition. Here, we suppose that the probability that \emph{collision} happens is $\varphi\in[0,1)$, whereas the probability that one of the conflicting pedestrians wins (and hence the target  cell  is assigned to him) is $1-\varphi\in(0,1]$. {In} the case where  \textcolor[rgb]{0,0,1}{a} collision does not happen, the winner is \emph{randomly} selected. If the pedestrian $n$ loses in one of \textcolor[rgb]{0,0,1}{the}  competitions, then he continues to compete for the next un-assigned preferred target cell until all the $k$ chances are used. If the pedestrians lose in all the $k$ rounds of competitions,  they do not change their locations  next time instant \footnote{\textcolor[rgb]{0,0,1}{According to our  observation and personal experience in the real world, when two pedestrians are  targeting the same location,    one is likely to change    his direction when the collision is about to happen.  This change of direction may happen almost at the same time when   another pedestrian  successfully comes to the targeted location. Therefore, in the rest of the paper, we suppose that each      pedestrian has one
chance to alter his target, i.e., $k=2$.}}.
\item Update the locations for all the pedestrians $n\in\mathbb{RM}(t)$. Since the pedestrians  {who reached}  one of the exits are understood as the ones who have successfully escaped from the evacuation system $\Pi$, release the escaped pedestrians from the simulation data set, let $t=t+1$, and update the set $\mathbb {RM}(t)$.
\item Calculate and output the pedestrian density $\rho_i(t)$ observed at the exit $i\in\mathcal N_{\rm ex}$.
\item Require the new information of $u_i(t)$, $i\in\mathcal N_{\rm EA}$, and go to (i) to continue until all of \textcolor[rgb]{0,0,1}{the}  pedestrians left the evacuation system $\Pi$ (i.e., $\mathbb {RM}=\emptyset$).

\end{enumerate}

\begin{rem}\emph{
It is important to note that the proposed evacuation model requires the inputs (i.e., the intensity coefficient $u_i$, $i\in\mathcal N_{\rm EA}$) from the guiding assistant system $\mathcal G$  and offers the real-time data (partial information) to the density controller for redesigning the inputs. Consequently, when a density controller is established, the  guiding signals are understood as the \emph{feedback} to the evacuation system for enhancing the efficiency of the evacuation process (see \autoref{fig:3}).}
 \end{rem}
%\begin{algorithm}[H]\label{Pau}
%\caption{Social force-based update rule of the characterized force-driven CA model}
%\KwIn{Intensity coefficients of guiding signals $u_k(t)$, $k\in\mathcal N_{\rm EA}$, from the guiding assistant system $\mathcal G$}
%\KwOut{The sampled partial information of the remaining pedestrians in the state space, e.g., crowd density around the exits }
%	(i) Calculate the social forces according to \autoref{social} and determine the sequence of ordered preference target cells  $P_n=[P_n^1,P_n^2,P_n^3,P_n^4]$  for all the remaining pedestrians  $n\in\mathbb {RM}(t)$
%
%    (ii)  Target competition.
%
%    (iii)	Define $\mathbb P(i)\triangleq\{ n\in \mathbb {RM}(k-1):P_n^i \neq 0\} $ and calculate the set $\mathbb S=\mathbb {RM}(k) \cap \mathbb P(i)$. $\mathbb S $ denotes the set of target cells have been occupied, and then we define the set $\mathbb {PT}= \mathbb P(i)- \mathbb S$.
%
%    (iv)	For each node $p$ in the set $\mathbb {PT}$, update $\mathbb {RM}(k)$ as follow:
%
%        	
%    \hspace{10pt}(a)Calculate the variable m which denotes the number of times that $p$ appears in $P_n$. Get the sequence $S(m)$  which denotes the indexes of $p$ appears in $P_n$.
%
%        	
%   \hspace{10pt} (b)If  $m>1$ then
%         Randomly select a pedestrian v from S(m) as the winner. And then put $\mathbb {RM}(k,v)=p$, $p_v^1=0$,$p_v^2=0$.
%       Otherwise put$\mathbb {RM}(k,S(m))=p$, $p_{S(m)}^1=0$,$p_{S(m)}^2=0$.
%
%	(5)Report$\mathbb {RM}(k)$ and the process stops.
%\end{algorithm}
%

\vspace{-7pt}
\subsection{Density Control Algorithm}\label{Sec23}
In this section, we consider an  on-off-based density control algorithm as the feedback control law for the guiding assistant system to adjust the guiding signals and hence affect the evacuation efficiency. Specifically, we suppose that all of the EAs only observe the values of pedestrian density $\rho_j(t)$ around their own assigned exit $j\in\mathcal N_{\rm ex}$ based on  some computer-aided technologies (e.g., image processing)  and use the observed information to update  $u_i$, $i\in\mathcal N_{\rm EA}$ \textcolor[rgb]{0,0,1}{for}  the guiding signals.
The density control algorithm considered in this paper  is   given by
{\setlength\abovedisplayskip{4pt}
\setlength\belowdisplayskip{4pt}
\begin{eqnarray}\label{BB1}
u_i(t)=\left\{\begin{array}{c}
                        1,\quad  \rho_j(t-\zeta)\leq\rho^{\rm aim}_i,\\
                        0,\quad  \rho_j(t-\zeta)>\rho^{\rm aim}_i,
                      \end{array}
\right. \quad i\in\mathcal N_{\rm EA},\quad t=1,2,3,\ldots,\vspace{-4pt}
\end{eqnarray}
where} $\zeta$ denotes the constant time step delay of the data (observed information), $j$ refers to the exit index assigned to \textcolor[rgb]{0,0,1}{the}  evacuation assistant $i$, and $\rho^{\rm aim}_i$ refers to the target pedestrian density set by the evacuation assistant $i$. In this case, the evacuation assistant $i$ turns on  (resp., turns off)  the guiding signal when the current density around its  exit is less (resp., larger) than the target density \footnote{If $\rho^{\rm aim}_i=1$ holds for   some $i\in\mathcal N_{\rm EA}$, then it is understood that those EAs never stop sending the guiding signals for attracting the pedestrian towards their own assigned exits. Alternatively, if $\rho^{\rm aim}_i=0$   holds for   some $i\in\mathcal N_{\rm EA}$, then it is understood that those EAs only begin to send the guiding signals when all the pedestrians around their assigned exits have already escaped from the evacuation system $\Pi$.}. %Here, we note that to enhance the evacuation efficiency, the selections of target  density $\rho^{\rm aim}_i$, $i\in\mathcal N_{\rm EA}$ are significantly important for the guiding assistant system $\mathcal G$.
\textcolor[rgb]{0,0,1}{It is important  to note that  when  $\rho^{\rm aim}_i=1$, $i\in\mathcal N_{\rm EA}$, the guiding signal is always opened and hence
 the guidance from the guiding assistant system $\mathcal G$ is understood as the classical (\emph{static}) guiding assistance. Alternatively, when  $\rho^{\rm aim}_i\not=1$, $i\in\mathcal N_{\rm EA}$,  it is understood that the classical (\emph{static}) guidance from the guiding assistant system $\mathcal G$ changes to  \emph{dynamic} guidance.  }

 \vspace{-3pt}
\begin{rem}\emph{
\textcolor[rgb]{0,0,1}{Note that a continuous-function control can be considered as the density control algorithm to enhance the evacuation efficiency if
the guiding signals $u_i$, $i\in\mathcal N_{\rm EA}$  are quantifiable. However, establishing quantifiable guiding
signals  relies on the structure (equipment) of the guidance system. For example, a PI control algorithm
can be used in the evacuation process when the strength of the guiding signals
is shown in the IoT devices to the pedestrians. Since the opening and closing of the guiding effect
do not rely on the structure of the guidance system, we use the on-off-based density control
in this paper.}}
\end{rem}\vspace{-6pt}

\emph{Problem}: Consider the guiding assistant system $\mathcal G$ with the density control algorithm shown in \autoref{BB1}. The main question investigated in this paper is how to set the observed regions and the target density  in the  guiding assistant system  to enhance  the evacuation efficiency as much as possible. Since the data collection (e.g., imaging processing) and data transmission  take time,
the  time step delay $\zeta$ may not be very small, and this parameter may affect the evacuation efficiency under the guiding assistant system $\mathcal G$. In the following sections, we first consider the case where the time step delay $\zeta$ is very small, and then we discuss the situation when  considering  a \textcolor[rgb]{0,0,1}{larger}   time step delay $\zeta$.

We evaluate the evacuation efficiency by using the nation of \emph{travel time} $T_{\rm end}$, which is defined as
the time instant when all the   pedestrians successfully escaped from the evacuation system, i.e., $\mathbb {RM}(t)=0$ at $t=T_{\rm end}$.
\textcolor[rgb]{0,0,1}{The dynamical guiding assistant system (or, the density control algorithm) is said to be effective if the  travel time is shorter  than the one under static guidance (i.e., $\rho^{\rm aim}_i=1$, $i\in\mathcal N_{\rm EA}$).}
%our main objectives in this paper are three folds: 1) Characterize the influence of the size of sampling regions on affecting evacuation efficiency through pedestrian density control 2) Characterize the influence of the initial number and distribution of pedestrians on affecting evacuation efficiency through pedestrian density control  3) Characterize the influence of the distribution of exits on affecting evacuation efficiency through pedestrian density control.

%Specifically, we suppose that all of the evacuation assistants observe the values of pedestrian density around their own location according to some computer-aided technologies, e.g., image processing.%

\vspace{-12pt}
\section{Simulation and Analysis}\label{section3}
\begin{figure}
  \centering
 \includegraphics[width=50mm]{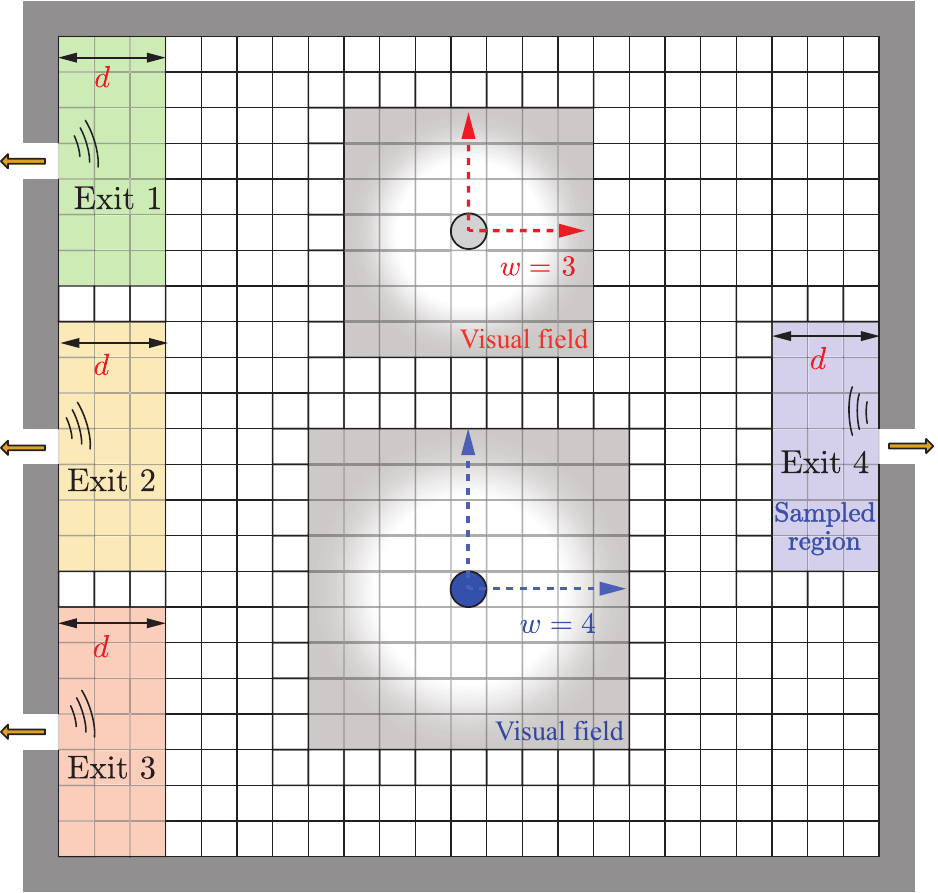}%step0_4
  \caption{Evacuation space with asymmetric exits layout   with $N=23$ and $\mathcal N_{\rm ex}=\mathcal N_{\rm EA}=\{1,2,3,4\}$, where the four colored regions around exits represent  the observed region for  calculating pedestrians' density with the radius being $d$. Two  pedestrians possessing the field of view of $w=3$ and $w=4$ are invisible to each other. Since none of the two pedestrians  appear in the observed regions, their information (location)  is unknown to the evacuation assistants. }
  \label{fig:5}
\end{figure}

\vspace{-4pt}
\textcolor[rgb]{0,0,1}{Limited by
the difficulty of obtaining real data under the COVID pandemic, the validity of the force-driven CA   model is
verified by the context of the fundamental results in \cite{chen2012study,gaofengqiang2016,REN2021125965}.} The main objective of this paper is to \textcolor[rgb]{0,0,1}{obtain  the optimal} control strategies
(e.g., target density, observed region) for the guiding assistant system $\mathcal G$ through analyzing simulation results.
In this section,  we  present and analyze our simulation results for an evacuation system $\Pi$ with $m=4$  exits and $\varepsilon=4$ evacuation assistants where every evacuation assistant is located \textcolor[rgb]{0,0,1}{at} the exit and is  observing the pedestrian density near its  exit to  turn on or shut down  the guiding signal according to \autoref{BB1}.
Specifically, letting   $N=23$, we  divide the indoor evacuation space to $23\times 23$ cells so that the evacuation system $\Pi$ may contain at most 529 pedestrians. We set $w_1=w_2=w_3=1$ in  \autoref{social}, $D=30$ in \autoref{guiding},   $E=60$ in \autoref{exit},  $\eta_1=10$, and $\eta_2=20$ in \autoref{mutual}. We consider the   evacuation space with   asymmetric exits layout as shown in \autoref{fig:5}.

 \vspace{-3pt}
\begin{rem}\emph{
\textcolor[rgb]{0,0,1}{In general, the effect of exit distribution (e.g., the number of exits and asymmetric degree of exits) may influence the evacuation efficiency (see the results in \cite{yue2011simulation}). For example, different from the main contribution of this paper, the effect of density control  with a symmetric exit distribution is revealed in \cite{REN2021125965} without considering  data time delay. In terms of the asymmetric exit distribution, it turns out that the tendency of simulation results shown in the following sections still can be similarly found when  we alter the distributions of those exits. Therefore, the evacuation space shown in \autoref{fig:5} is considered without loss of generality in the rest  of this paper. }}
\end{rem}\vspace{-6pt}

For simplicity, we let $\rho^{\rm aim}_i=\rho_{\rm aim}$ for all $i\in\mathcal N_{\rm ex}$ and define the observed region as a rectangular area expanded from the exit $i\in\mathcal N_{\rm ex}$ with $d\in\mathbb N_+$ steps (see the typical example of the four observed regions with $d=3$ in \autoref{fig:5}, where each of the EAs is able to observe at most 21 pedestrians near its exits).
Recalling the fact that the time step delay $\zeta$ of the observed data depends on the data collection structure, we first  assume that the time step delay $\zeta$ is very small in \autoref{small} and then relax this assumption in \autoref{large} to discuss how the result changes when we have a more significant  time step delay $\zeta$.
It is important to note that as one of the pedestrians' inherent properties, the size of the \textcolor[rgb]{0,0,1}{visual field} may be  diverse from each other and hence the pedestrians existing in the evacuation system $\Pi$ may be   homogeneous  or heterogeneous.
As the simulation result might be strongly connected to the mentioned inherent property (i.e., size of the \textcolor[rgb]{0,0,1}{visual field}),
to derive a strategic conclusion on how to set the target density and observed region  without loss of generality, we may consider the evacuation processes   with heterogeneous pedestrians \textcolor[rgb]{0,0,1}{in} the following sections.

%
% During the experiment, we will adjust the size of the sampling regions, the distribution and initial number of pedestrians $N_p$,  the value of target density $\rho_{\rm i,aim}$ for exit $i\in\mathcal N$,  and the distribution of exits,  so as to reveal the influence of the factors which affect pedestrian density control on evacuation,  where $w_{sample}$ is ranged from 3 to 7, $\rho_{\rm i,aim}$  is ranged from 0.1 to 0.8,  and $N_p$ is ranged from 125 to 500.  From  ,  we can see that the capacity of the sampling regions is $w_{sample}$,  hence we define the pedestrian density near exit $i$ at the time instant $k$ as $\rho_i(k)={\rm num}_i(k)/(w_{sample}\times7)$,  where ${\rm num}_i(k)$ denotes the number of pedestrians located in the region $i$.  Note that these parameters described later are fixed in the experiment.
%\vspace{-6pt}

%\subsection{Homogeneous Pedestrians With The Same Size of Field of View}
%In the following sections, we illustrate  our simulation results for the situation considering  homogeneous pedestrians with the same size of   field of views.
%
%We first show the necessity of using dynamic guiding assistances, and then we characterize the influences of target density and the size of the observed regions on the evacuation efficiency.
%\vspace{-6pt}

\subsection{Small Time Step Delay in Density Control Algorithm}\label{small}
\subsubsection{Necessity of Using Dynamic Guiding Assistances}\label{exmaple}
\begin{figure}
  \centering
  \includegraphics[width=100mm]{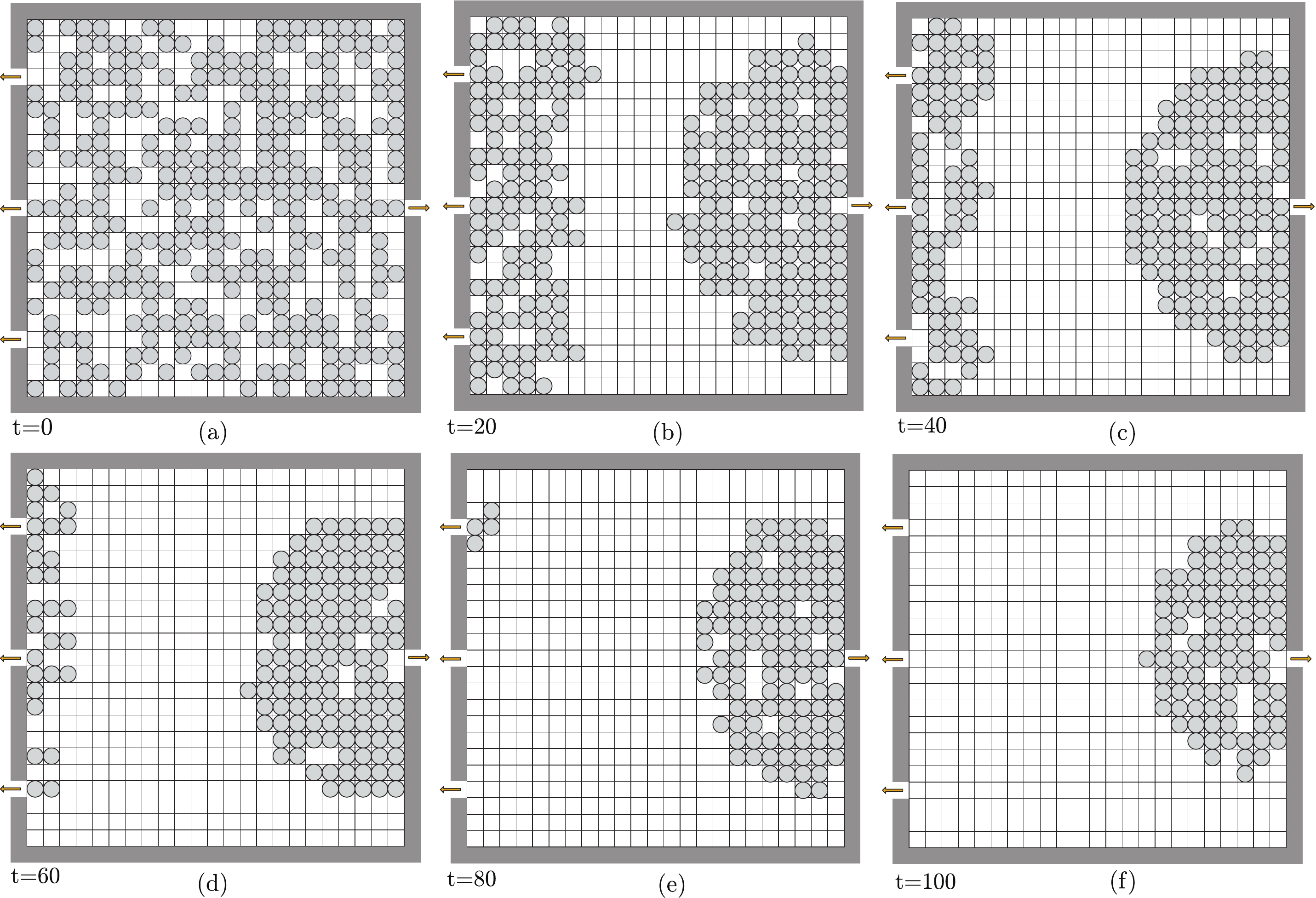}
  \caption{Distributions of homogeneous pedestrians under \emph{static} guiding assistance. Then guiding signal is always opened in the evacuation process since $\rho_{\rm aim}=1$. Pedestrians immediately huddle near the exits and never change the moving direction to the reverse direction. A large group of pedestrians huddle around one of the exits  (exit 4) even  though the other 3 exits are not utilized at all,  which makes the evacuation process run in an inefficient way. }
  \label{fig:02}
\end{figure}

  In the beginning, to indicate the necessity of using dynamic guiding assistance in the asymmetric-exits evacuation process,
we use the initial distribution   shown in \autoref{fig:02}(a) in the simulation where 317 pedestrians are  distributed randomly in the evacuation space. The field of view is given by $w=3$ for all the pedestrians so that each one of them is able to observe at most 48  pedestrians around its  location (see the grey circle in  \autoref{fig:5} representing a pedestrian with $w=3$).
Letting the collision probability be  $\varphi=40\%$, {the distributions of the pedestrians  under \emph{static} guiding assistance (i.e., $\rho_{\rm aim}=1.0$)  are shown in \autoref{fig:02}.}
 It can be seen from the figure that since the EAs never stop attracting the pedestrians, whatever how large the density is in its observed region, the pedestrians are suddenly   separated in the horizontal direction, and hence a blank zone appears in the middle of the evacuation space from the time instant $t=20$. The blank zone expands along with  time  and the utilization of the three exits in the left-hand side of the evacuation space is  almost vacant  in the later stage of the evacuation, i.e., the remaining pedestrians \emph{only}  locate around   exit 4 after time instant $t=60$.
As a result, it is observed that a lot of evacuation efficiency is wasted in the evacuation process under  the static  guiding assistance.

\begin{figure}
  \centering
  \includegraphics[width=100mm]{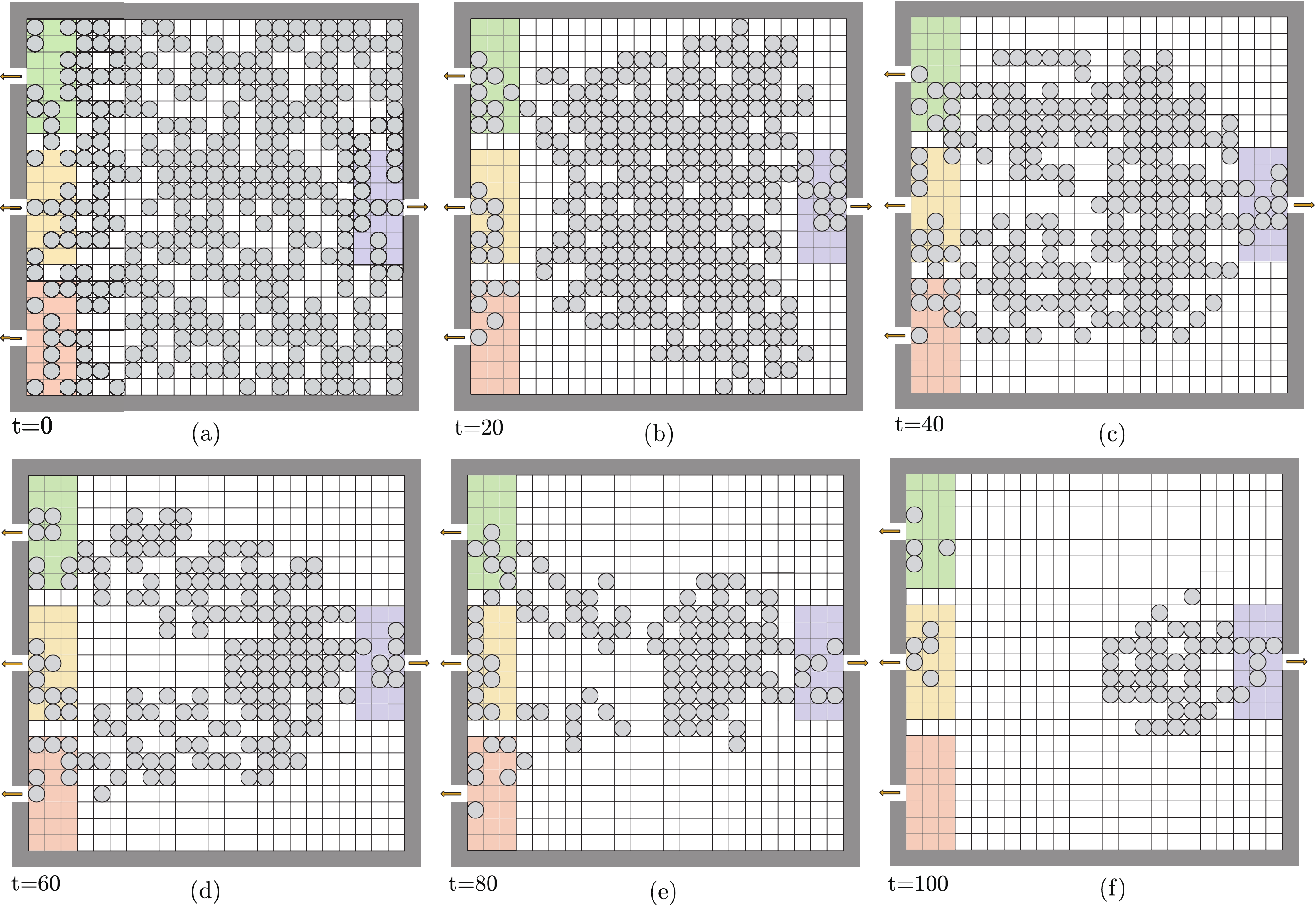}
  \caption{Distributions of homogeneous pedestrians under \emph{dynamic} guiding assistance with $\rho_{\rm aim}=0.3$ and $d=3$, $\zeta=1$. The 4 colored regions represent the regions at which the evacuation assistants observe the crowd density. The evacuation assistant turns on or shuts down the guiding signal according to the real-time data of the observed crowd density.  Pedestrians do not huddle near the exits but stay in the evacuation space's central area until the pedestrians in the observed regions are all evacuated.   The 4 exits are almost utilized efficiently even at the time instant $t=100$. Compared  to the simulation results under static guidance in  \autoref{fig:02},  this example indicates the significance of using \emph{dynamic} guiding assistance in the asymmetric-exits evacuation process.}
  \label{fig:0}
\end{figure}

 Next, we  show a typical example of  pedestrian distributions, which reveals the effectiveness and significance of using \emph{dynamic} guiding
assistance. Specifically, in the simulation, we let $\rho_{\rm aim}=0.3$, $\zeta=1$, $d=3$ (see the colored areas representing the  observed regions in \autoref{fig:5}). {The distributions of the pedestrians  are  illustrated in \autoref{fig:0}.}  Compared  to the case with \emph{static} guiding assistance, the amounts of the remaining  pedestrians at the time instants $t=20$, $40$, $60$, $80$, and $100$  under \emph{dynamic} guiding assistance in \autoref{fig:0} are \emph{certainly} less than the ones in \autoref{fig:02}. It can be easily found in \autoref{fig:0} that the pedestrian densities in the four  observed regions under control are much less than the ones in  \autoref{fig:02} so that the possibility  of collisions and congestions happening around the exits is suppressed, which make the evacuation run more efficiently. Another direct  and interesting observation found in the comparison is that the pedestrians are no longer separated into two crowds in the horizontal direction but stay in the central region under  \emph{dynamic} guiding
assistance (see  \autoref{fig:0}(b)--(e)).
{Those simulation results}  capture the fact that the   dynamic guiding assistant system with only partially observable data may still positively control (affect) the location of all   pedestrians  in the multi-exit
evacuation process by involving the density control algorithm.

\begin{figure}
  \centering
  \includegraphics[width=120mm]{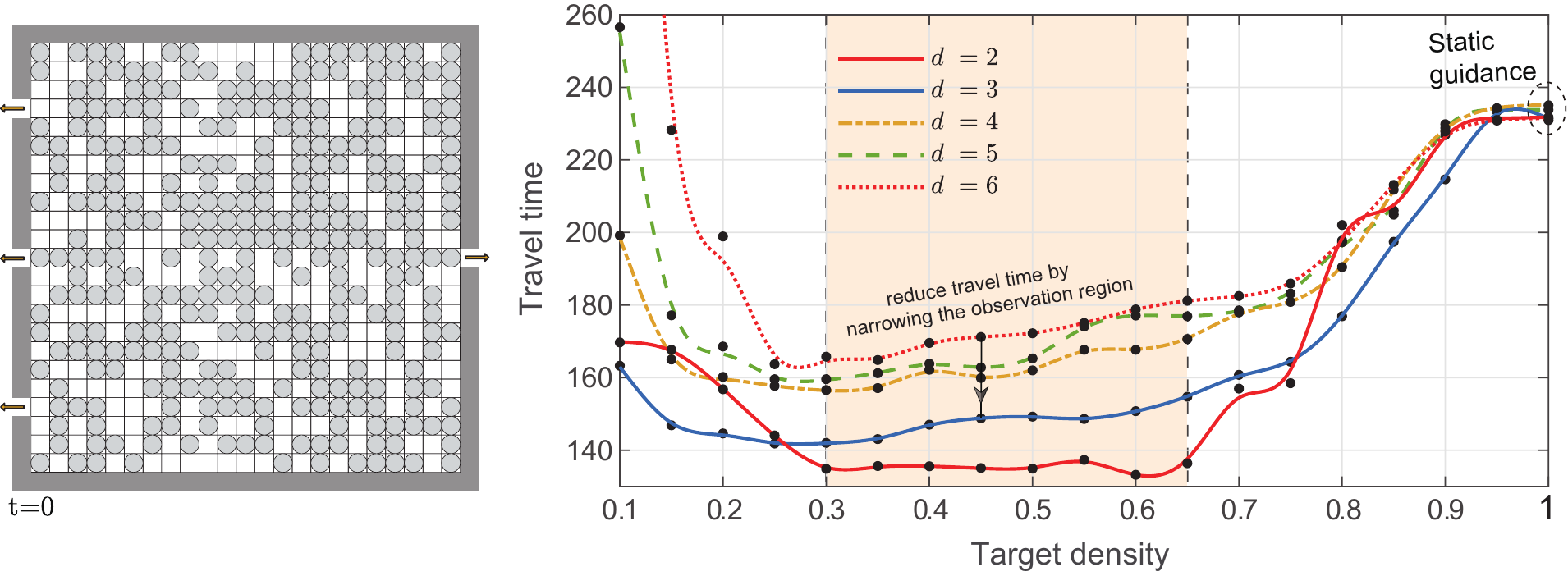}
  \caption{Travel time versus the target density  $\rho_{\rm aim}$ in density control when $\zeta=1$ with different sizes   of the observed regions. To avoid  misinterpretation  caused by the stochastic process existing in the evacuation model, all the data of average travel times are generated from more than 30   simulations.  The initial distribution   is shown on  the left-hand side of the figure.   \textcolor[rgb]{0,0,1}{The fitting
results are generated by the ``Smoothing Spline'' function of the Curve Fitting box in Matlab
with the goodness of fit (R-square) being larger than 0.995.}
The optimal size of the observed region is found as 2. That is to say, the observed region is suggested to be small for the case where the time step delay $\zeta$ is small.  The suggested target density for the dynamic evacuation assistant system ranges  from 0.3 to 0.65.
 }
  \label{fig:16}
\end{figure}

\begin{figure}
  \centering
  \includegraphics[width=100mm]{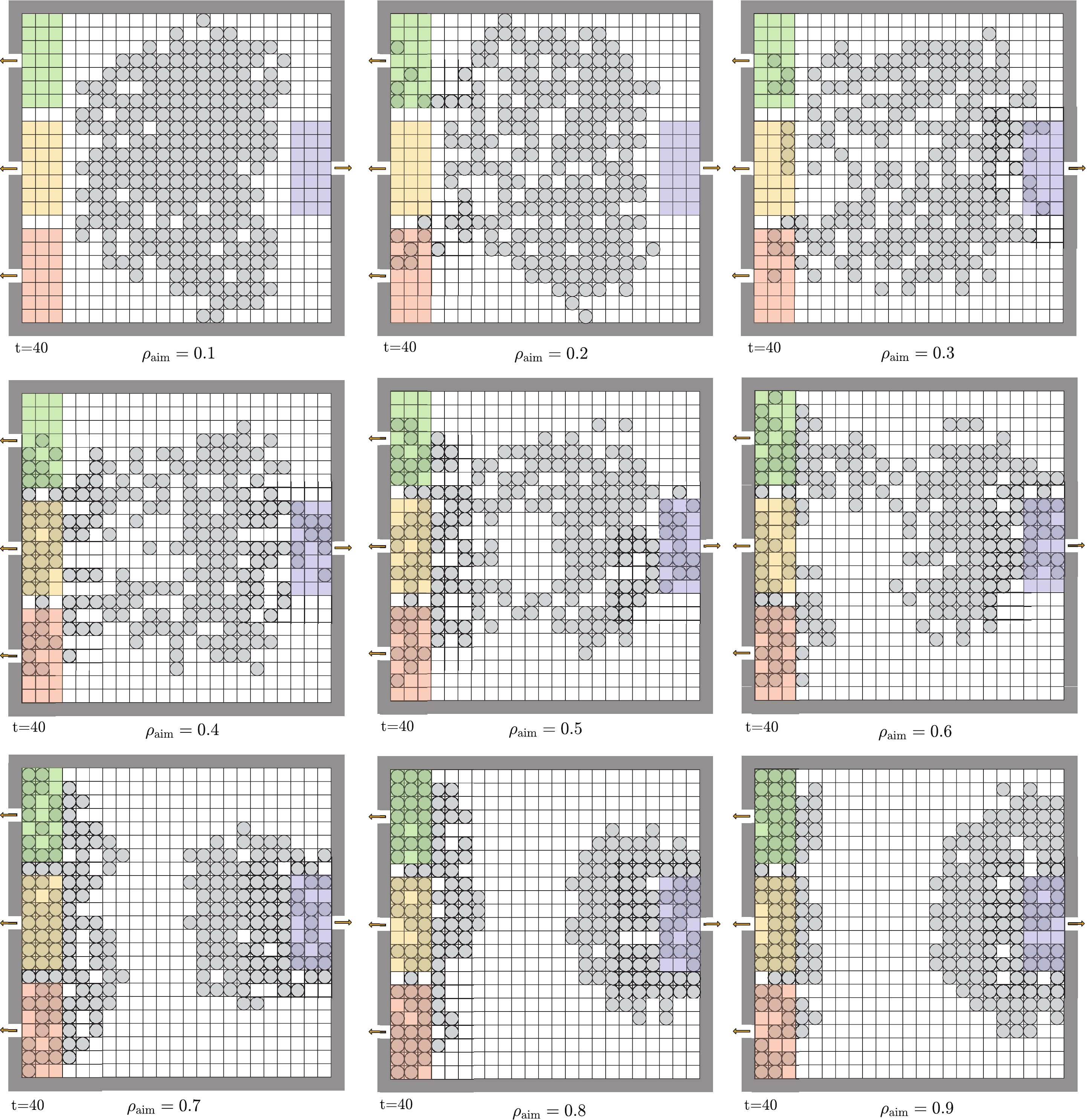}\vspace{-8pt}
  \caption{Distributions of homogeneous pedestrians under \emph{dynamic} guiding assistance  with   different target densities at the  time instant $t = 40$ when $\zeta=1$,  $d=6$. It is observed that the pedestrians are slowly attracted towards the exits when the target density is small, and an arching phenomenon occurs when the target density is large.
The pedestrians are obviously separated in the horizontal direction for $\rho_{\rm aim}\geq 0.7$, which should be avoided as  discussed in \autoref{exmaple}. The number of pedestrians trapped in the arching phenomenon increases when the target density increases.
 }
  \label{fig:40}\vspace{-13pt}
\end{figure}

\subsubsection{Influence of Target Density and Size of Observed Regions}\label{Sec312}
In this section, we  investigate the influence of  the target density of the dynamic guiding assistant system on the travel time of the evacuation process    and hence derive the optimal size of the observed regions. Letting $\rho_{\rm aim}\in[0.1,1]$, the  response curves of   travel time of the evacuation process on target density are shown in \autoref{fig:16}  with several different sizes of the observed regions  where the    collision probability is still set to $\varphi=40\%$.  Note that the presented curves are yielded from the average  value of the data from more than 30 times   simulations and show similar tendencies with respect to $\rho_{\rm aim}$. On  the one hand, it can be seen from the figure that compared  to the case with static guidance  (i.e., when target density equals 1.0), the evacuation efficiency is significantly improved  when the target density $\rho_{\rm aim}$ is moderate in   the density control algorithm  \autoref{BB1}.
 Without loss of generality, letting the size of observed regions be $d=3$, \autoref{fig:40} shows  the  pedestrian distributions at the  time instant $t = 40$ when different target densities are used in the simulations. This  figure provides   evidence to support that the target density for the dynamic evacuation assistant system should be a moderate value. In particular, it is observed from \autoref{fig:40} that  pedestrians are slowly attracted towards the exits when the target density is small, and \textcolor[rgb]{0,0,1}{an} arching phenomenon occurs when the target density is large\footnote{Arching phenomenon refers to the scene in which a large group of
pedestrians huddle around the exit in the evacuation process. This phenomenon  leads to serious conflicts and congestions among the pedestrians, and eventually reduce the traffic efficiency of the exits. During designing the dynamic guiding assistant system, the arching phenomenon should be suppressed as much as possible to avoid conflicts and congestions.}.
On the other hand,  it can be further seen from \autoref{fig:16} that except for the case with $d=2$, the efficiency improvement made by the  density control algorithm   drops off when the size of the observed regions increases. The optimal  size $d$ for the observed regions and the optimal target density $\rho_{\rm aim}$ are found as $d=2$ and $\rho_{\rm aim}\in[0.3,0.7]$, respectively.
Note that this observation reveals an interesting fact that to enhance the evacuation efficiency, we only need to observe the pedestrians' location from a small region near the exit instead of a large region when the time step delay in the density control algorithm  (\ref{BB1}) is very small.
%the effective target density that improves the evacuation efficiency is getting narrower and narrower .   On the other hand, it can be seen  from \autoref{fig:16} that the improvements of evacuation efficiency made by the dynamic guiding assistances with the size $d_i=3$, $i\in\mathcal N_{\rm EA}$ are lager than the ones in $d_i=5$, $i\in\mathcal N_{\rm EA}$, $d_i=4$, $i\in\mathcal N_{\rm EA}$, and $d_i=6$, $i\in\mathcal N_{\rm EA}$.
%Moreover,  However,
%because   the  travel time under $d_i=2$, $i\in\mathcal N_{\rm EA}$, is larger than the one under $d_i=3$, $i\in\mathcal N_{\rm EA}$ for some of the target density, it is observed that the evacuation efficiency under dynamic guiding assistances with $d_i=2$, $i\in\mathcal N_{\rm EA}$ is better than  the one with $d_i=3$, $i\in\mathcal N_{\rm EA}$ only for the set of target density $\rho_{\rm aim}\in[0.3,0.7]$.

\begin{figure}
  \centering
  \includegraphics[width=85mm]{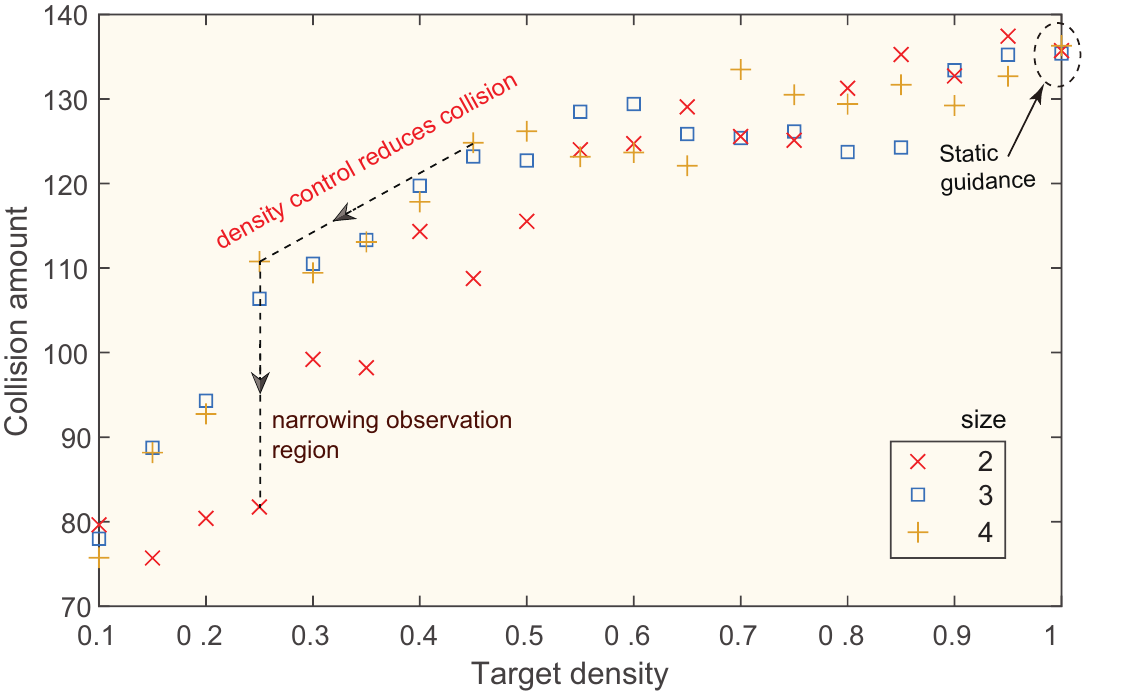}\vspace{-11pt}
  \caption{ \textcolor[rgb]{0,0,1}{Amount of collisions (that trigged around the exits) versus  target density  with different sizes of the observation region  when $\zeta=1$. The   collision  amount   is  reduced  when we narrow the observation
region for the dynamic guiding assistant system. Decreasing the target density of the on-off-based density control can significantly reduce the collisions around the exits.  }}
  \label{fig:18}
\end{figure}

\begin{figure}
  \centering
  \includegraphics[width=92mm]{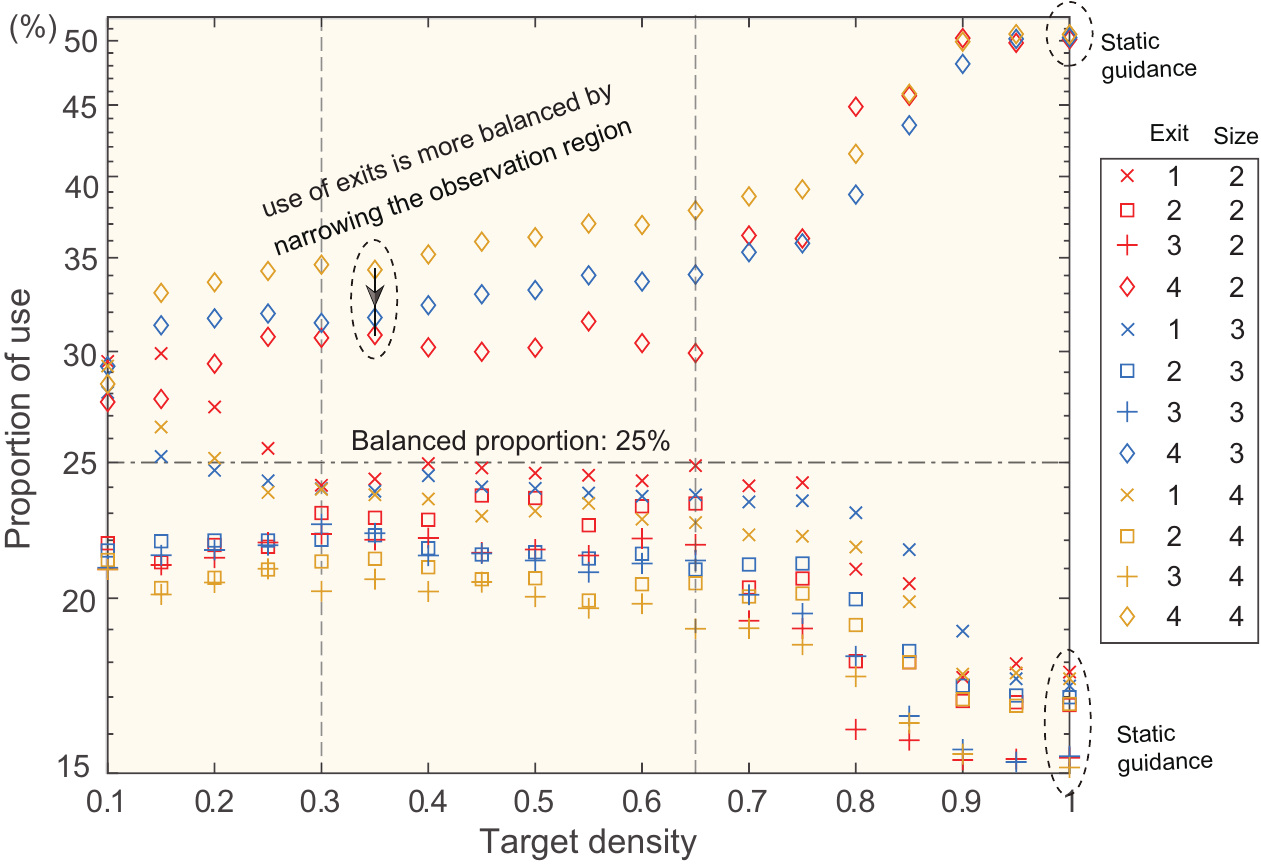}
  \caption{ \textcolor[rgb]{0,0,1}{Proportion of  use of each exit  versus  target density  with different sizes  of the observation region  when $\zeta=1$. The use of exits is more balanced by
narrowing the observation region for the dynamic guiding assistant system. The  use of exits  is extremely imbalanced under  static guidance.}}
  \label{fig:17}
\end{figure}

 To further understand the intrinsic reason why the  target density and the size of the observed region matter in the evacuation process, we  illustrate the response curves of collision amounts   (that trigged around the exits)  on the target density $\rho_{\rm aim}$ in \autoref{fig:18}.
 It can be observed from the figure that  the
collision amount is reduced when we narrow the observation region for the dynamic guiding assistant system. Moreover, it can be seen that decreasing the target density of the
on-off-based density control (\ref{BB1}) can significantly reduce the collisions around the exits in the evacuation process.
   On the other hand, the proportion of use of each exit versus target density with different sizes  of the observation region is demonstrated in \autoref{fig:17}. It can be seen from this figure that   the use of exits is extremely imbalanced under   static
guidance and the density control can significantly balance the use of the four asymmetric exits, especially when the target density is set to $\rho_{\rm aim}\in(0.3,0.7)$. Also, it can be observed  that the use of exits is  more
balanced by narrowing the observation region for the dynamic guiding assistant system.

\begin{figure}
  \centering
  \includegraphics[width=130mm]{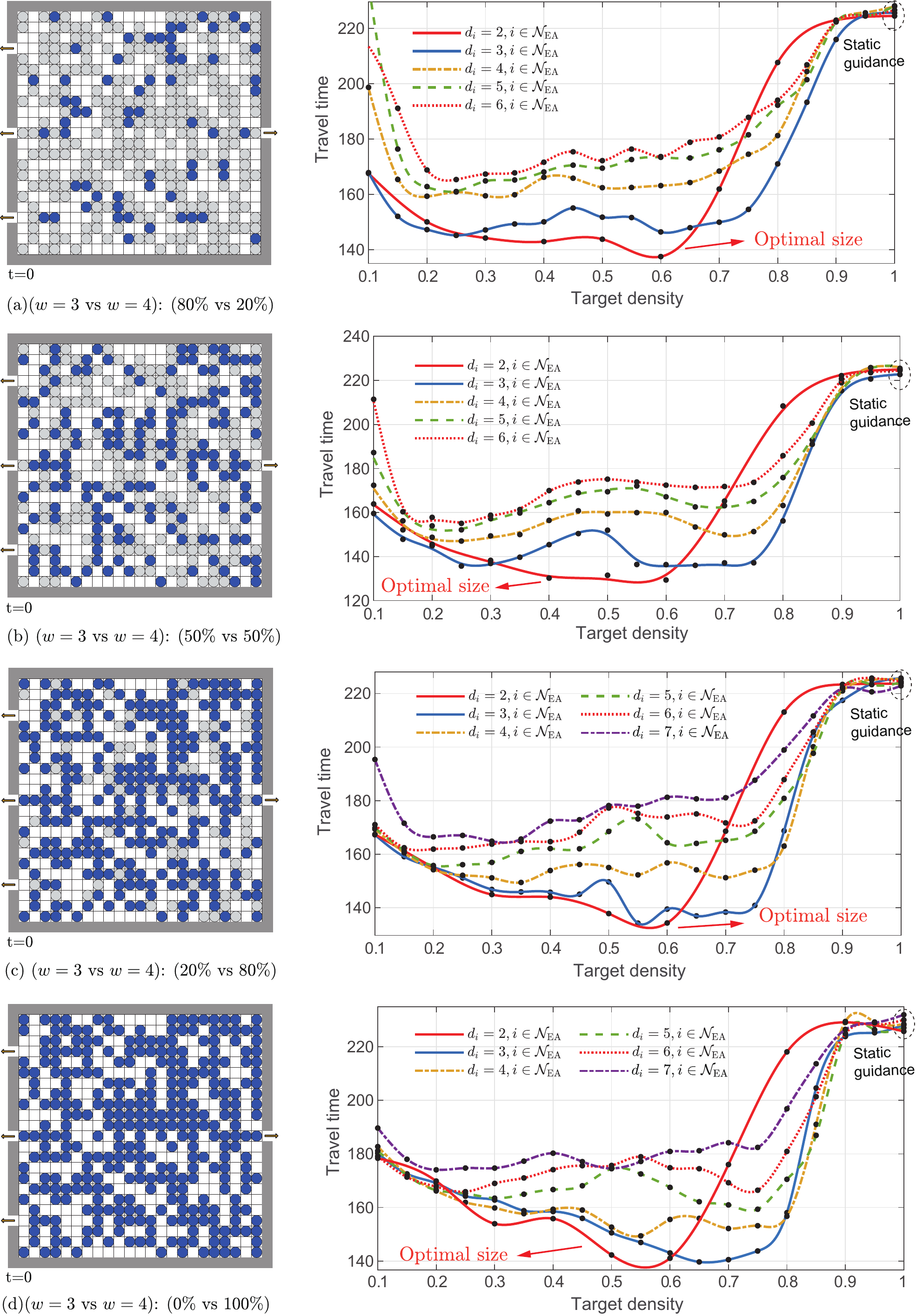}
  \caption{Response of  the target density on travel time under density control with mixed or pure pedestrians possessing different or the same view sizes when $\zeta=1$. (a) 20 percent of the   pedestrians possess a bigger \textcolor[rgb]{0,0,1}{visual field} in the simulation. (b) 50 percent of    pedestrians possess a bigger \textcolor[rgb]{0,0,1}{visual field} in the simulation. (c) 80 percent of    pedestrians possess a bigger \textcolor[rgb]{0,0,1}{visual field} in the simulation. (d) All    pedestrians possess a bigger \textcolor[rgb]{0,0,1}{visual field} in the simulation. The initial distributions set in the simulations are shown
the left-hand side  of the sub-figures where the blue circles denote  the pedestrians with a larger \textcolor[rgb]{0,0,1}{visual field} (i.e., $w=4$ instead of $w=3$). \textcolor[rgb]{0,0,1}{All of the fitting
results are generated by the ``Smoothing Spline'' function of the Curve Fitting box in Matlab
with the goodness of fit (R-square) being larger than 0.995.} In all of the cases,  the optimal size of the observed region is found as 2.  That is to say, the observed  region is suggested to set as a small one around the exit for the case with small time step delay $\zeta$.   }
  \label{fig:38}
\end{figure}

 Now, it is important to note that the above observations of the optimal size $d$ are without loss of generality when we consider  different  initial distributions or sizes of the  visual field.
 For example, considering the situation where the pedestrians in the initial time instant may be mixed by some taller pedestrians  possessing a larger field of view,
we present  the simulation results with heterogeneous pedestrians holding the \textcolor[rgb]{0,0,1}{visual field} with $w=3$ and $w=4$  under different   mixing ratios   in \autoref{fig:38}.  Even though the tendencies of the curves are slightly different from \autoref{fig:16}, we note that the size $d=2$ remains  the optimal size for the observed regions in all of the simulations under different  mixing ratios. Moreover, when $d\geq 3$, it is still observed that  the efficiency improvement made by the
density control algorithm drops off when the size of the observed regions increases for each of the sub-figures.
%The curves optimal   target densities are diverse in \autoref{fig:16}, but the simulation result under $\rho_{\rm aim}=0.5$ shows a quite low travel time for $d=2$ comparing to the other target densities and sizes of the observed regions.
%Recalling the fact that  the size of the field of view is the  pedestrian's inherent property, we note that the pedestrians in the real evacuation may possess different view size. For example, . Hence, it is necessary to consider the situation with   sizes of the field of view.
%In this section, we mix the pedestrians with the view size being $w=3$ with the (taller) pedestrians with the view size being $w=4$ in the simulation and investigate whether the optimal size of the observed regions derived in previous section changes or not. In all of the simulations under different mixed percentage shown in , it is found that when the size of the observed region is bigger than 2, the evacuation efficiency is possible to be further enhanced by minimizing the size of the observed regions.
Even though the  mixing ratios change the tendency of the response curves, the travel time under the target density $\rho_{\rm aim}\in(0.5,0.6)$ with $d=2$ is always the minimum value.   As a result, we conclude that  no matter how much the mixing ratio  is in the simulation,
the observed  region is suggested to be small for the case where the time step delay $\zeta$ is small.

\subsection{Influence of Data Time Delay}\label{large}
\begin{figure}
  \centering
  \includegraphics[width=100mm]{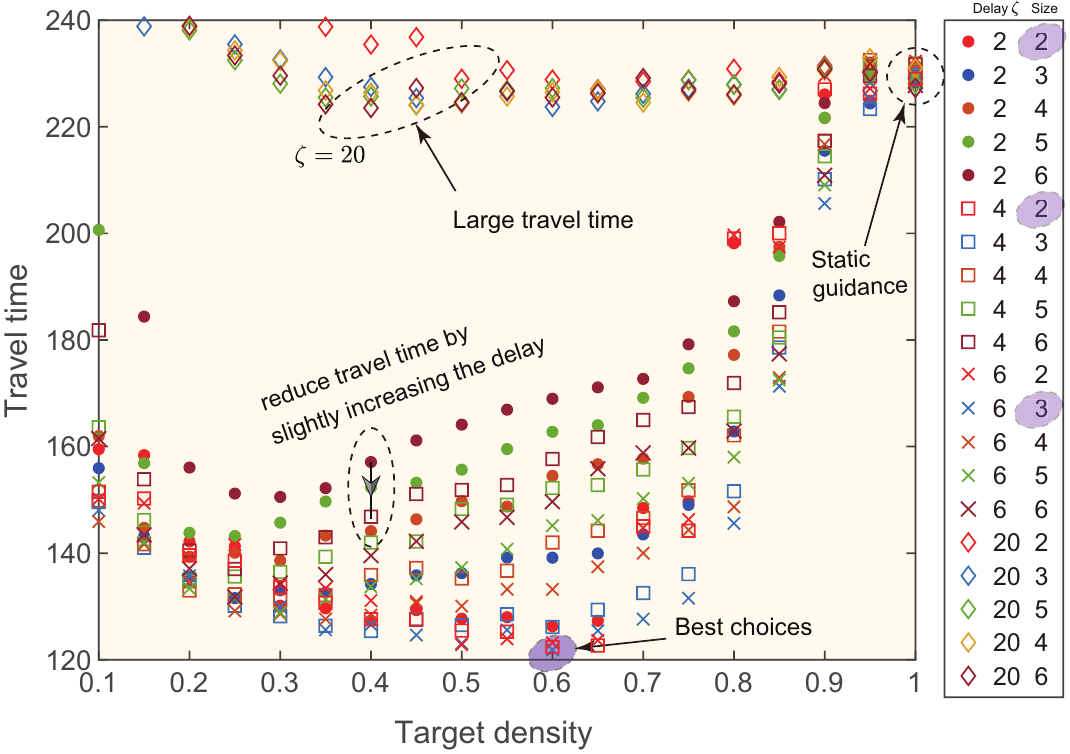}
  \caption{Travel time versus the target density  $\rho_{\rm aim}$ in density control with different sizes   of the observed regions and different time delay $\zeta$.  The initial distribution considered in the simulations is the same one that we used in \autoref{fig:16}. \textcolor[rgb]{0,0,1}{The travel time is reduced by
slightly increasing the delay.  However, when the time delay $\zeta$ is too large,  the density control algorithm may no longer  enhance the evacuation efficiency. } }
  \label{fig:48}
\end{figure}
Recalling the fact that the time step delay $\zeta$ of the density control algorithm is a given positive parameter depending on the data collection structure of the guiding assistant system, it is important to discuss how the optimal size changes when the given time step delay $\zeta$ changes. The travel time versus the target density  $\rho_{\rm aim}$ in density control with different sizes   of the observed region and  time delays is shown in \autoref{fig:48}, where we used \autoref{fig:16} as the initial distribution in all the simulations. It can be seen from \autoref{fig:48} that the travel time can be reduced by
slightly increasing the delay.  When the given time step delay $\zeta$ is larger, setting up a larger region to observe the pedestrians' location is better.
Those observations are natural  in the sense of physical meaning, since the data with a large delay are not able to represent the characteristics of the pedestrians' current situation for a small observed region, but for a large  observed region,  those data may  be  able to represent the partial characteristics of the pedestrians' real situation because of  the intrinsic inertia.
However, when the time delay $\zeta$ is too large,  the density control algorithm may no longer obviously enhance the evacuation efficiency (see \autoref{fig:48}). As a result, the time step delay allowed in our proposed density control algorithm is bounded. The above numerical findings give important insights on designing  computer-aided (control-based) guiding strategies in real evacuations.

\section{Conclusion}\label{section:4}
%Different from the existing works in terms of guided emergency evacuation which only consider a  single exit evacuation scenario based on a social force model.
%For example, Yang et al. \cite{yang2016necessity} first claimed the necessity of guides in pedestrian  .
%Ma et al. \cite{ma2016effective,ma2017dual} considered dynamic leaders inside a room who attract the crowds and move together with them towards a unique exit with different  pedestrian densities.
%The po ssibility and effects of controlling the pedestrian densities were not considered yet in the literature.

To enhance the evacuation efficiency in partially observable asymmetric-exit evacuation \textcolor[rgb]{0,0,1}{under guidance, a general framework of the dynamic guiding assistant system
was proposed to investigate the  effect of density control.}   In the characterized system,
 multiple evacuation assistants  are established to observe the partial information of pedestrians'
location and adjust the guiding signals of the dynamic guiding assistant systems according to the observed information (i.e., pedestrian densities in the observed regions near the corresponding exits). Specifically,  a  simple on-off-based density control algorithm associated with a target density is considered for    evacuation assistants based on the delayed data of  (observed) pedestrian densities to meet the physical  challenges on data collection,  transmission, and implementation  which  often exist in the realistic computer-aided evacuation process.
By involving a force-driven CA model, we presented the simulation results in accordance with data delays to eventually give a strategic suggestion on how to set the observed region and the target density.

According to our numerical simulations, we first revealed the necessity of using dynamic guiding assistance in the asymmetric-exit evacuation process.
It was found that the proposed density control algorithm can control (positively affect) the global distribution of the pedestrians' locations and suppress arching phenomena in the evacuation process even using  partially observed information under time delays. To derive the optimal size of the observed regions, we investigated the influence of the target density of the dynamic guiding assistant system on the
travel time of the evacuation process.
It was found that the dynamic guiding assistant system with
only partially observable data can suppress the triggers of collisions around the exits and avoid  inefficiently separating the pedestrians  in the evacuation process. After showing various simulation results, we revealed an interesting fact without loss of generality that to enhance the evacuation efficiency, we only need to observe
the pedestrians' location from a small region near the exit instead of a \textcolor[rgb]{0,0,1}{large} region when the time step delay in the
density control algorithm is very small. Furthermore, we found that the time step delay allowed in our proposed
density control algorithm is bounded since it may be impossible to significantly enhance the evacuation efficiency by using the density control algorithm when the  time step delay is too large. For both  the small delay case and large delay case, we suggested  the target density of the density control algorithm to be   a moderate value.  Our numerical results
are expected to provide  insights into designing  the computer-aided guiding strategies in  real evacuations.

\textcolor[rgb]{0,0,1}{However, due to  the difficulty of inviting
human volunteers under the COVID pandemic, conducting an  actual experiment is expected
in  future. Moreover, there may be many
different heterogeneities in the evacuation process, such as the psychological considerations
(risk-averse/risk-seeking), quality of the pedestrians, will on sharing visual field, to name but a few. The analysis on those factors may be promised necessary future research directions.}
As a preliminary study connecting the control theory and pedestrian evacuation theory, we only used a simple and intuitive control algorithm for  the evacuation assistant system. However, there may exist  possible future research directions when we apply some novel control algorithms to the evacuation process, such as model predictive control \cite{kohler2020computationally}, system identification \cite{mauroy2019koopman}, Q-learning for optimal control \cite{lee2018primal}, noise elimination, etc.

\section*{CRediT Authorship Contribution Statement}
{\bf Fengqiang Gao}: Investigation; formal analysis; writing- original draft preparation, reviewing (equal).
{\bf Zhihao Chen}: Software; validation;  reviewing (equal).
{\bf Yuyue Yan}: Conceptualization; methodology; investigation; writing- review $\&$ editing (lead).
{\bf Linxiao Zheng}: Visualization; reviewing (equal).
{\bf Huan Ren}: Group management; reviewing (equal).
\section*{Acknowledgments}

This work was supported jointly by Program for Young Excellent Talents in University of Fujian Province (201847) and China Scholarship Council (201908050058).
We thank  Xie Chen for participating in a discussion in the early state of this work.



%% References

\bibliographystyle{elsarticle-num}
\bibliography{autosam}






% \end{thebibliography}
%\appendix
%
% \section{Detailed expression of social forces}

\end{document}